\documentclass[]{amsart}

\usepackage{amsmath}
\usepackage{amssymb}
\usepackage{color}
\usepackage{natbib}
\usepackage{epsfig}
\usepackage{subfigure}

\newtheorem{theo}{Theorem}
\newtheorem{lemma}{Lemma}
\newcommand{\SF}{{\mathcal F}}
\newcommand{\N}{{\mathbb N}}
\renewcommand{\P}{{\mathbb P}}
\newcommand{\Ess}[3]%
   {\operatorname{E}_{#1}\left[\left.#2\vphantom{#3}\right|#3\right]}
\newcommand{\E}{\operatorname{E}}

\title[Efficient swaptions in Hull-White]%
{Efficient swaptions price in Hull-White one factor model}

\author[M. Henrard]%
   {Marc Henrard}

\date{First version: 20 December 2008; this version: 9 January 2009}

\address{Marc Henrard \\ Head of Interest Interest Rate Modelling\\ Dexia Bank \\ Belgium}

\email{Marc.Henrard@dexia.com}

\keywords{swaption, Hull-White one factor model, approximation, calibration}

\thanks{JEL classification: G13, E43, C63}

\thanks{AMS mathematics subject classification: 
  91B28, 91B24, 91B70, 60G15, 65C05, 65C30}

\begin{document}

\begin{abstract}
The Hull-White one factor model is used to price interest rate options. The parameters of the model are often calibrated to simple liquid instruments, in particular European swaptions. It is therefore very important to have very efficient pricing formula for simple instruments. Such a formula is proposed here for European swaption. Based on a very efficient corrector type approximation the approximation is efficient both in term of precision and in term of spped. In our implementation the approximation is more than ten time faster than the direct pricing formula and more than twenty time faster than the Jamshidian trick.
\end{abstract}

\maketitle

\section{Introduction}

Due to its flexibility, the \cite{HUW.1990.1} one factor model is often use as a simple and efficient model to price interest rate derivatives, including exotics.

The model parameters are often obtained through a calibration procedure. A calibration is a reverse engineering work were the model parameters are reconstructed from market prices. In most of the cases this inversion is not obtained explicitly but implicitly by solving numerically a non-linear equation. The price function being the one to inverse. In the numerical procedure, the pricing function is used repeatedly. This is the reason why a numerically efficient price computation of vanilla instruments is very important.

In this note the price of European swaption in the Hull-White one factor model is discussed. The first exact pricing solution proposed for that model is probably the one proposed by \cite{JAM.1989.1}. Its solution is based on a decomposition, now called \emph{Jamshidian's trick}. The decomposition consists in dividing the bond (or swap) in a set of zero-coupon bonds with strikes such that all of them are exercised in the same conditions. As the model is a one-factor model with prices monotonous in the underlying stochastic variable, such a level exists. In the Jamshidian decomposition, the underlying random variable is the short-rate. A one dimensional equation consisting of exponentials with relatively complex coefficients needs to be solved to find the (common) exercise level. The pricing is obtained by adding zero-coupon options. For a $n$ coupon bond, there are $2*n$ normal cumulative functions to compute.

A more direct solution was proposed in \cite{HEN.2003.1}. Using an explicit description of the zero-coupon price in term of a normally distributed random variable, the swaption price is written as the discounted cash-flow multiplied by some weights (probabilities). The exercise level in term of the random variable is computed by solving a one dimensional equation. The equation is similar to the one in the Jamshidian decomposition; it consists of exponentials but its coefficients are more easy to write. The price is obtained with one normal cumulative functions for each coupon plus one for the strike. For a $n$ coupon bond, there are $n+1$ such functions to compute. Depending of the implementation, this second method is usually two to four times faster than the Jamshidian method.

\section{Model, hypothesis and preliminary results}

The HJM framework describes the behavior of $P(t,u)$, 
the price in $t$ of the zero-coupon bond
paying $1$ in $u$ ($0\leq t, u \leq T$).
When the discount curve $P(t,.)$ is absolutely continuous, which is 
something that is always the case in practice as the curve is constructed by
some kind of interpolation, there exists $f(t,u)$ such that 
\begin{equation}
\label{EqnP}
P(t,u) = \exp\left( -\int_t^u f(t,s) ds \right).
\end{equation}
The idea of \cite{HJM.1992.1} was to exploit this 
property  by modeling $f$ with a stochastic differential equation  
\[
df(t,u) = \mu(t,u) dt + \sigma(t,u) dW_t
\]
for some suitable (potentially stochastic) $\mu$ and $\sigma$ and deducing the 
behavior of $P$ from there. To ensure the arbitrage-free property of
the model, a relationship between the drift and the volatility is 
required. The model technical details can be found in the 
original paper or in the chapter 
\emph{Dynamical term structure model} of \cite{HUK.2004.1}.  

The probability space is $(\Omega, \{\SF_t\}, \SF, \P)$. The 
filtration $\SF_t$ is the (augmented) filtration of a one-dimensional
standard Brownian motion $(W_t)_{0\leq t \leq T}$.
To simplify the writing in the rest of the paper, the notation
\[
\nu(t,u) = \int_t^u \sigma(t,s) ds
\]
is used.

Let $N_t = \exp(\int_0^t r_s ds)$ be the cash-account numeraire with
$(r_s)_{0\leq s \leq T}$ the short rate given by $r_t = f(t,t)$. The
equations of the model in the numeraire measure associated to $N_t$ are
\[
df(t,u) = \sigma(t,u) \nu(t,u) dt + \sigma(t,u) dW_t
\]
or
\[
dP^N(t,u) = - P^N(t,u) \nu(t,u) dW_t
\]

The notation $P^N(t,s)$ designates the numeraire rebased value of $P$, i.e.\ 
$P^N(t,s) = N_t^{-1} P(t,s)$.

The following \emph{separability hypothesis} will be used:

\begin{description}
\item[H] The function $\sigma$ satisfies $\sigma(t,u) = g(t) h(u)$ 
for some positive functions $g$ and $h$.
\end{description}

Note that this condition is essentially equivalent to the condition (H2) of
\cite{HEN.2003.1} but written on $\sigma$ instead of on $\nu$. The condition 
on $\nu$ was $\nu(s,t_2)-\nu(s,t_1) = f(t_1,t_2) g(s)$.

We recall the 
generic pricing theorem \cite[Theorem 7.33-7.34]{HUK.2004.1}.

\begin{theo}
\label{ThPrGen}
Let $V_T$ be some $\SF_T$-measurable random variable.  If $V_T$
is attainable, then the time-$t$ value of the derivative is given by 
$V_t^N = V_0^N + \int_0^t \phi_s dP_s^N$ where $\phi_t$ is the strategy and
\[
V_t = N_t \Ess{\N}{V_T N_T^{-1}}{\SF_t}.
\]
\end{theo}

We now state two technical lemmas that were presented
in \cite{HEN.2005.1}.
 
\begin{lemma}
\label{Lem1}
Let $0\leq t\leq u \leq v$.  
In a HJM one factor model, the price of the zero coupon bond
can be written has,
\[
P(u,v) = 
\frac{P(t,v)}{P(t,u)} \exp \left( -\frac12 \int_t^u 
\left(\nu^2(s,v) - \nu^2(s,u) \right) ds + \int_t^u 
\left(\nu(s,v)-\nu(s,u)\right) dW_s \right).
\]
\end{lemma}

\begin{lemma}
\label{Lemr}
In the HJM one factor model, we have
\[
N_u N_v^{-1} = \exp\left(-\int_u^v r_s ds\right) = 
P(u,v) \exp\left( \int_u^v \nu(s,v) dW_s - 
\frac12 \int_u^v \nu^2(s,v)ds \right).
\]
\end{lemma}

We give the pricing formula for swaptions for a future time 
(see \cite[Theorem 2]{HEN.2006.1}). The notations used to describe the swaption are the following. Let $\theta$ be the expiry date and the swap is represented by its cash-flow equivalent $(t_{i}, c_{i})_{i=0,\ldots,n}$. The date $t_{0}$ is the swap start date and $t_{i}$ ($i=1,\ldots,n$) are the fix coupon 
dates. The amounts $c_{0}$ is $-1$\footnote{It is $-K$ for a bond option of strike $K$.}, $c_{i}>0$ ($i=1,\ldots,n-1$) are the coupons and $c_{n}>0$ is
the final coupon plus 1 for the notional.

The swap is described by its cash-flow equivalent $(c_i,t_i)$ ($0\leq i \leq n$). The dates $t_{i}$ are the settlement date $t_{0}$, the (fixed) coupon payment dates $t_i$ ($1\leq i \leq n$). In the descriptions a receiver swaption is considered. The values are the initial notional $c_0=-1$ (resp. strike $c_0=-K$), the different coupons $c_i$ and the final cash-flow including the coupon and the notional. Usually the coupons are regular with $c_i = \delta_i R_K$ for a strike rate $R_K$ and a day count fraction $\delta_i$ between $t_{i}$ and $t_{i+1}$. The swaption expiry is denoted $\theta$ and $\theta\leq t_{0}$.

\begin{theo}
\label{ThSwpt}
Suppose we work in the HJM one-factor model with a volatility term of the form 
(H2).  Let $\theta \leq t_0 < \cdots < t_n$, $c_0<0$ and $c_i\geq 0$ 
($1\leq i\leq n$).  The price of an European receiver swaption, 
with expiry $\theta$ on a swap with cash-flows $c_i$ and cash-flow dates
$t_i$ is given at time $t$ by the $\SF_t$-measurable random variable
\[
\sum_{i=0}^n c_i P(t,t_i) N(\kappa + \alpha_i)
\]
where $\kappa$ is the $\SF_t$-measurable random variable defined as the
(unique) solution of 
\begin{equation}
\label{EqnKappa}
\sum_{i=0}^n c_i P(t,t_i) \exp\left( -\frac12 {\alpha_i}^2 - 
\alpha_i \kappa \right) = 0
\end{equation}
and
\[
{\alpha_i}^2 = \int_t^\theta \left(\nu(s,t_i)-\nu(s,\theta)\right)^2 ds.
\]
The price of the payer swaption is
\[
- \sum_{i=0}^n c_i P(t,t_i) N(-\kappa - \alpha_i)
\]
\end{theo}

The Brownian motion change between the $N_t$ and the 
$P(t,t_{j+1})$ numeraires is given by
\[
dW_t^{j+1} = dW_t + \nu(t, t_{j+1}) dt.
\]

\section{Swaption pricing}

From an option pricing point of view, swaptions and bond options are equivalent. From now on we will refer only to swaption. Up to the change of names, all the results are valid both for swaptions and bond's options.

The forward values of the zero-coupon bonds and swap without the initial notional are given in $t$ by
\[
P^i_t = \frac{P(t,t_i)}{P(t,t_0)}
 \quad \mbox{and} \quad 
B_t = \frac{\sum_{i=1}^n c_i P(t,t_i)}{P(t,t_0)}.
\]
Those value are the value rebased by the numeraire $P(.,t_0)$

Let $\nu^i(t) = (\nu(t,t_i)-\nu(t,t_0))$ ($0\leq i\leq n$). 
In the martingale probability associated to the numeraire $P(.,t_0)$, the rebased prices are martingale that satisfy the equations
\[
dP^i_t = - P^i_t \nu^i(t) \cdot dW_t^0.
\]
The zero-coupon bond prices are exactly log-normal as the volatility $\nu^i(t)$ is deterministic. 

The rebased swap value satisfy
\[
dB_t = -\sum_{i=1}^n c_i P_t^i \nu^i(t) \cdot dW_t^0.
\]
Using the notation $\alpha_t^i = c_i P_t^i/B_t$ and $\sigma(t) = \sum \alpha^i_t \nu^i(t)$, the equation is becoming
\[
dB_t = - B_t \sigma(t) \cdot dW_t^0.
\]
This is formally a log-normal equation but the $\sigma$ coefficient is state dependent.

The volatility $\sigma(t)$ of the coupon bond can be approximated by its initial value
\[
\alpha_t^i \simeq \alpha_0^i = c_i \frac{P_0^i}{B_0}, \quad \sigma_0(t) = \sum_{i=1}^n \alpha_0^i \nu^i(t).
\]

Let 
\[
\bar{\sigma_0}^2 = \int_0^\theta \left| \sigma_0(t) \right|^2 dt.
\]
With the approximation the equation for the bond $B$ is log-normal and the standard Black formula approach can be used:
\[
R_0 = P(0,t_0) \E^0\left[ P^{-1}(\theta, t_0) \left(B_\theta - K P(\theta,t_0) \right)^+ \right] \simeq P(0,t_0) \E^0 \left[ \left( B_0 \exp\left(\bar{\sigma_0} X - \frac12 \bar{\sigma_0}^2 \right) - K \right)^+ \right]
\]

\begin{theo}[Initial freeze approximation]
In the HJM model, the price, with initial freeze approximation, of a receiver swaption with expiry $\theta$ is given at time 0 by
\[
R_0 = P(0,t_0) \left( B_0 N(\kappa+\bar{\sigma_0}) - K N(\kappa) \right)
\]
where
\[
\kappa = \frac{1}{\bar{\sigma_0}} \left(\ln\left(B_0/K\right) - \frac12 \bar{\sigma_0}^2\right).
\]
The price of a payer swaption is
\[
P_0 = P(0,t_0) \left(K N(-\kappa) -  B_0 N(-\kappa-\bar{\sigma_0}) \right)
\]
\end{theo}

This formula is equivalent to pricing the bond with the Black formula and an implied volatility equal to $\bar{\sigma_0}$.

\cite{BAV.2006.1} proved that this approximation is very efficient in the context of the BMM. Intuitively the largest part of the swap is in the notional at maturity $t_{n}$. That part is exactly log-normal and the total is well approximated by a log-normal dynamic. In the example analyzed in the relevant section it will showed that this is true also for amortized swaptions for which the notional is paid by in trenches along the swap's life.

\subsection{Corrector style approximation}

The predictor-corrector method is often used in Monte Carlo simulations linked to the LMM. It was initially developed by \cite{KLP.1995.1} and used in the framework of the Libor Market Model by \cite{HJJ.2001.1}. It consists for a given path to first simulate a step with the Euler scheme (equivalent to the initial freeze approximation). The value obtained is used as an approximation of the step end state (predictor). Then the (state dependent) coefficient is approximated by the average of the coefficient at the starting state and the coefficient at the approximated end state (corrector). This approach leads to very efficient Monte Carlo implementations.

The above described method is implemented at the path level. For each path the approximated final value is different and the state dependent coefficients approximation is different. The path level approximation is suitable for Monte Carlo implementation but not for approximated explicit formulas. It is possible to design a method that works with the same idea of averaging coefficients, not at the path level but at a more general level. The standard approach works at the path level and is the same for all options priced with the same Monte Carlo. The new approach described here extends the one proposed in \cite{HEN.2007.1} and is based on approximation adapted to the option priced and valid for all paths.

The approach is first described in an intuitive and loose way before going through the details.
In the Black-Scholes formula the option price can be viewed as today's prices of the option components ($B_0$, $K$) weighted by some probabilities. If the probabilities are well approximated, the price is good. The main probability used is the one of reaching the strike from the current position. In the Black-Scholes formula it is roughly $N(d_{1})$. For the paths that reach the strike the coefficients can be approximated in the corrector way. The corrector state is given by a strike state. The corrected volatility is the average between the initial coefficient and the strike (final) one. In our case the figure represents the coefficient of a log-normal distribution that ends up on the strike for the same paths as the real distribution. Said in another way, this is the (approximative) equivalent log-normal volatility for which the probability to reach the strike is equal to the actual one. In options term, it is the \emph{implied volatility} that gives the same price for that strike.

The value of the different rates and zero-coupon bond prices at the strike are not unique as the model is multi-factors. The exact way the strike information is transformed into a coefficient information need to be decided. There is one constraint for $m$ parameters. The choice is within a $m\!-\!1$-manifold into a $m$ dimensional space: an infinity of potentially large dimension but with zero-measure.

The description of the option price in term of probability is not exact. The option is a probability average of the pay-off. The price obtained through the approach described above will not be exact. It is a significant improvement to the initial freeze/Euler approach.

A \emph{strike} value of the different parameters has to be selected. The swap price is \emph{at-the-money} at expiry when
\[
\sum_{i=0}^n c_i P^i_\theta = 0
\]
The discounting value of the zero-coupon bond can be approximated (initial freeze) by
\[
P^i_\theta = \frac{P(0,t_i)}{P(0,t_0)} \exp\left(-\tau_i X_i - \frac12 \tau_i^2\right)
\]
with
\[
\tau_i^2 = \int_0^\theta \left| \nu_0^i(t) \right|^2 ds
\]
and the $X_{i}$ standard normally distributed random variables.
By choosing (arbitrarily) to have all the stochastic variables $X_i$ equal at the strike the (one dimensional) equation to solve is
\[
\sum_{i=0}^n c_i P_0^i \exp\left(-\tau_i \bar{x} - \frac12 \tau_i^2\right) = 0.
\]
Obtaining the solution to the above equation requires to solve a one dimensional equation equivalent to the one solved in the swaption price for the Gaussian HJM model (\cite{HEN.2003.1}). For numerical reasons one may prefers not to have to solve this type of equation. The above equation can be replaced by it first order approximation
\[
\sum_{i=0}^n c_i P_0^i \left(1-\tau_i \bar{x} - \frac12 \tau_i^2\right),
\]
the solution of which is explicit:
\[
\bar x = \frac{\sum_{i=0}^n c_i P_0^i- \frac12 \sum_{i=0}^n c_i P_0^i\tau_i^2}
{\sum_{i=0}^n c_i P_0^i\tau_i}.
\]

The zero-coupon prices, in the exponential case and the approximated first order case, are given by
\[
P_K^i = P_0^i \exp\left(-\tau_i \bar{x} - \frac12 \tau_i^2\right), \quad \mbox{respectively} \quad 
P_K^i = P_0^i \left(1 -\tau_i \bar{x} - \frac12 \tau_i^2\right).
\]
The rates and bond prices are
\[
\prod _{j=0}^{i-1} (1+\delta_j L_K^j) = (P_K^i)^{-1} \quad \mbox{and} \quad B_K = \sum_{i=1}^n c_i P_K^i = K.
\]

By defining 
\[
\alpha_K^i = \frac{c_i P_K^i}{B_K}
\]
the swaption can be priced with a (option strike dependent) approximated volatility
\begin{equation}
\label{EqnSigmaK}
\bar{\sigma}_K(t) = \frac12 \left( \sum_{i=1}^n (\alpha_0^i + \alpha_K^i) \nu^i(t) \right).
\end{equation}

Note that in the multi-factor model, the volatility $\bar{\sigma}_K(t)$ is a vector, as $\nu_{0}$ and $\nu_{K}$ are.

\begin{theo}[Corrector implied volatility]
\label{ThCor}
In the local volatility LMM, the price, with initial freeze and corrector approximation, of a receiver swaption is given at time 0 by
\[
R_0 = P(0,t_0) \left( B_0 N(\kappa_K + \bar{\sigma}_K) - K N(\kappa_K) \right)
\]
where
\[
\kappa_K = \frac{1}{\bar{\sigma}_K} \left(\ln\left(B_0/K\right) - \frac12 \bar{\sigma}_K^2\right)
\]
and 
\[
\bar{\sigma}_K^2 = \int_0^\theta \left| \bar{\sigma}_K(t) \right|^2 dt.
\]
The price of a payer swaption is
\[
P_0 = P(0,t_0) \left(K N(-\kappa_K) -  B_0 N(-\kappa_K-\bar{\sigma}_K) \right)
\]
\end{theo}

\subsection{Hull-White volatility}

The different coefficients used in the previous sections are explicitly described for the Hull-White one factor model.

In the case of the extended Vasicek or one-factor Hull and White model, one has $\sigma(s,t) = \eta \exp(-a(t-s))$ and $\nu(s,t) =  (1-\exp(-a(t-s))) \eta/a$ in the constant volatility case. The time-dependent volatility is also covered with $\sigma(s,t) = \eta(s) \exp(-a(t-s))$ and $\nu(s,t) =  (1-\exp(-a(t-s))) \eta(s)/a$. The $\alpha$ used in Theorem~\ref{ThSwpt} are given in the constant volatility case by
\[
\alpha_{i}^2 = \frac{\eta^2}{2a^3} \left(\exp(-a\theta)-\exp(-at_{i})\right)^2 
\left(\exp(2a\theta)-1)\right).
\]

In the time-dependent case, $\eta$ is piece-wise constant with $\eta(s)=\eta_i$ for $s_{i-1}\leq s \leq s_{i}$ and $0=s_0<s_1<\cdots < s_n=+\infty$. The expiry dates are between some of those dates and the relevant dates are denoted $s_{q-1}  < \theta \leq s_q$. To shorten the notation an intermediary notation is used: $r_0 = s_0 < r_l = s_l < r_q = \theta$. With those notations, one has
\[
\alpha_{i}^2 = \frac{1}{2a^3} \left(\exp(-a\theta)-\exp(-at_{i})\right)^2 
\sum_{l=0}^{q-1} \eta_l^2 \left( \exp(2ar_{l+1})-\exp(2ar_l) \right).
\]

The zero-coupon bond instantaneous volatility $\nu^i(t)$ is given by
\[
\nu^i(t) = \frac{1}{a} \eta(s) \exp(at) (\exp(-at_0)-\exp(-at_i)).
\]
The $[0,\theta]$-period volatility is similar to $\alpha_i$ and is given by
\[
\tau_i^2 = \frac{1}{2a^3} \left(\exp(-at_0)-\exp(-at_{i})\right)^2 
\sum_{l=0}^{q-1} \eta_l^2 \left( \exp(2ar_{l+1})-\exp(2ar_l) \right).
\]

From there all the coefficients used in the pricing follow from simple arithmetic operations.

\section{Approximation numerical analysis and comparison}

\subsection{Approximation quality}

The approximation quality is estimated in the following way. An extended Vasicek model with time volatilities is calibrated to market data. The data are the yield curve and the at-the-money volatilities. This is done with two different date and to types of calibration to ensure some diversity in the test.

Once the model is calibrated, it is used to price a wide range of swaptions using both the exact pricing formula and the approximated formula. The Black swaption implied volatility of the results is used as comparison figure. In term of precision, the market quote the Black volatilities with a maximum precision of 0.10\%. Any approximation error below this can be consider as below the measure error. Nevertheless we a priori require a error below 0.025\% for options up to 300bps out-of-the money. It is compared to other approximations proposed in the literature.

The first example is done with market data from 4 July 2008. The model is calibrated to diagonal ATM options with expiry between 1 and 10 years and tenors between 10 and 1 year. The mean reversion parameter used is 0.02.

\begin{figure}[ht]
\begin{center}
\subfigure[Volatility]%
{\epsfig{file=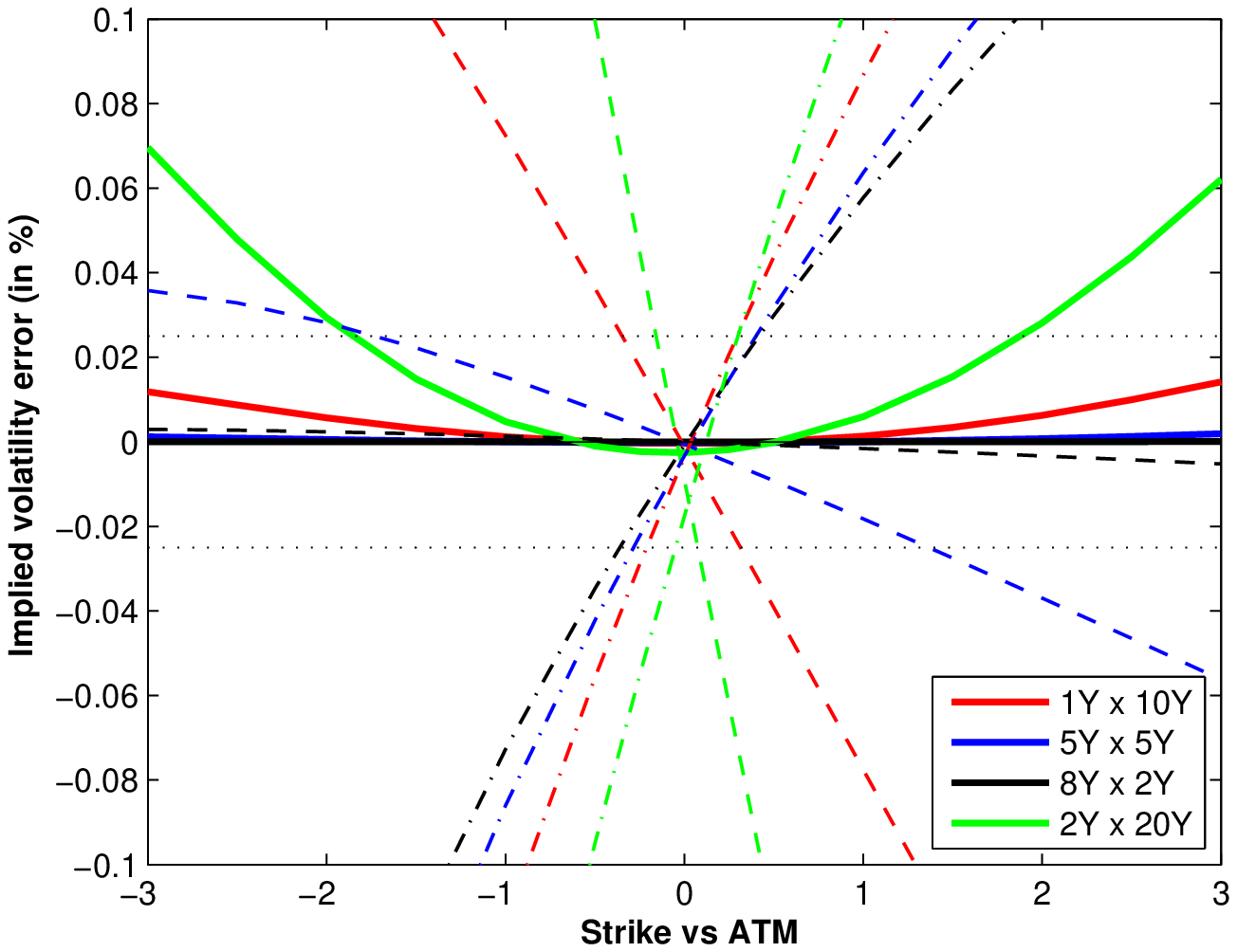, height=0.30 \textheight,width=0.48 \textwidth}
\label{FigErr1V} }
\hfill
\subfigure[Price]%
{\epsfig{file=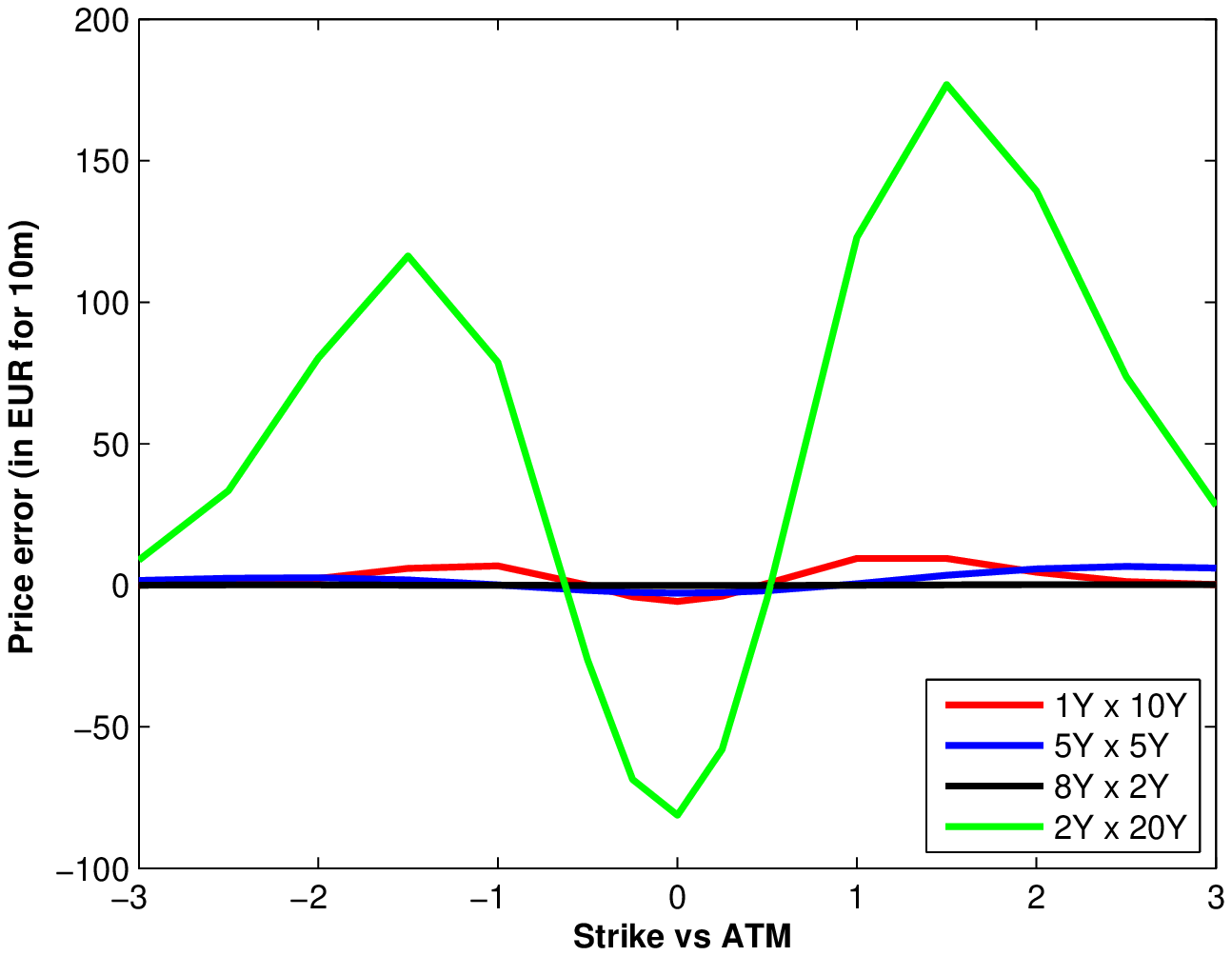, height=0.30 \textheight,width=0.48 \textwidth}
\label{FigErr1P} }
\end{center}
\caption{Error introduced by the approximation. Market data as of 4 July 2008.}
\label{FigErr1}
\end{figure}

With that calibration, four type of swaptions are priced: 1Y x 10Y, 5Y x 5Y, 8Y x 2Y and 2Y x 20Y. For each of those types, a large range of strikes, up to 300 bps away from the money is used. The results are given in Figure~\ref{FigErr1}.

The Figure contains the error for the proposed approximation and the initial freeze approximation. The improvement of the method is clear on the picture. For all the tested swaps, except for the longest maturity and the most extreme strikes, the error is within the a priori band.  Note that the options for which the error is above the a priori limit the strike is 300 bps away from the money; that level of moneyness is usually not quoted in the market. For most of the errors, it is difficult to distinguish them from 0 in the graph. For the most extreme case, where the error is above the a priori limit, it may be useful to look at the error in term of price instead than in term of volatility. For deep out-of-the-money option an error on the volatility will have a smaller impact on the price (Figure~\ref{FigErr1P}). In that case it appear clearly that even if the extreme strikes have higher volatility error, the price error is relatively small.

\begin{figure}[ht]
\begin{center}
\subfigure[Volatility]%
{\epsfig{file=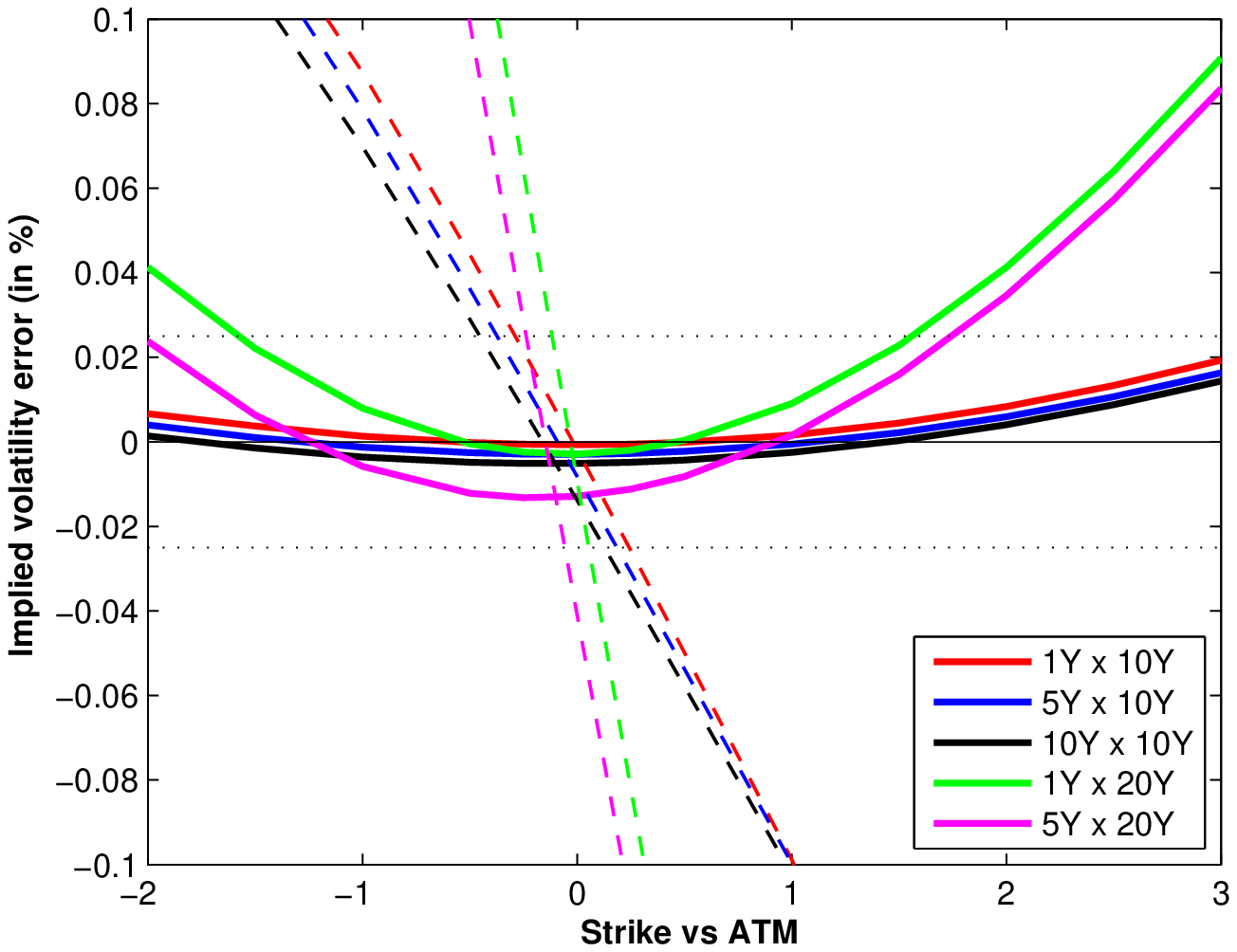, height=0.30 \textheight,width=0.48 \textwidth}
\label{FigErr2V} }
\hfill
\subfigure[Price]%
{\epsfig{file=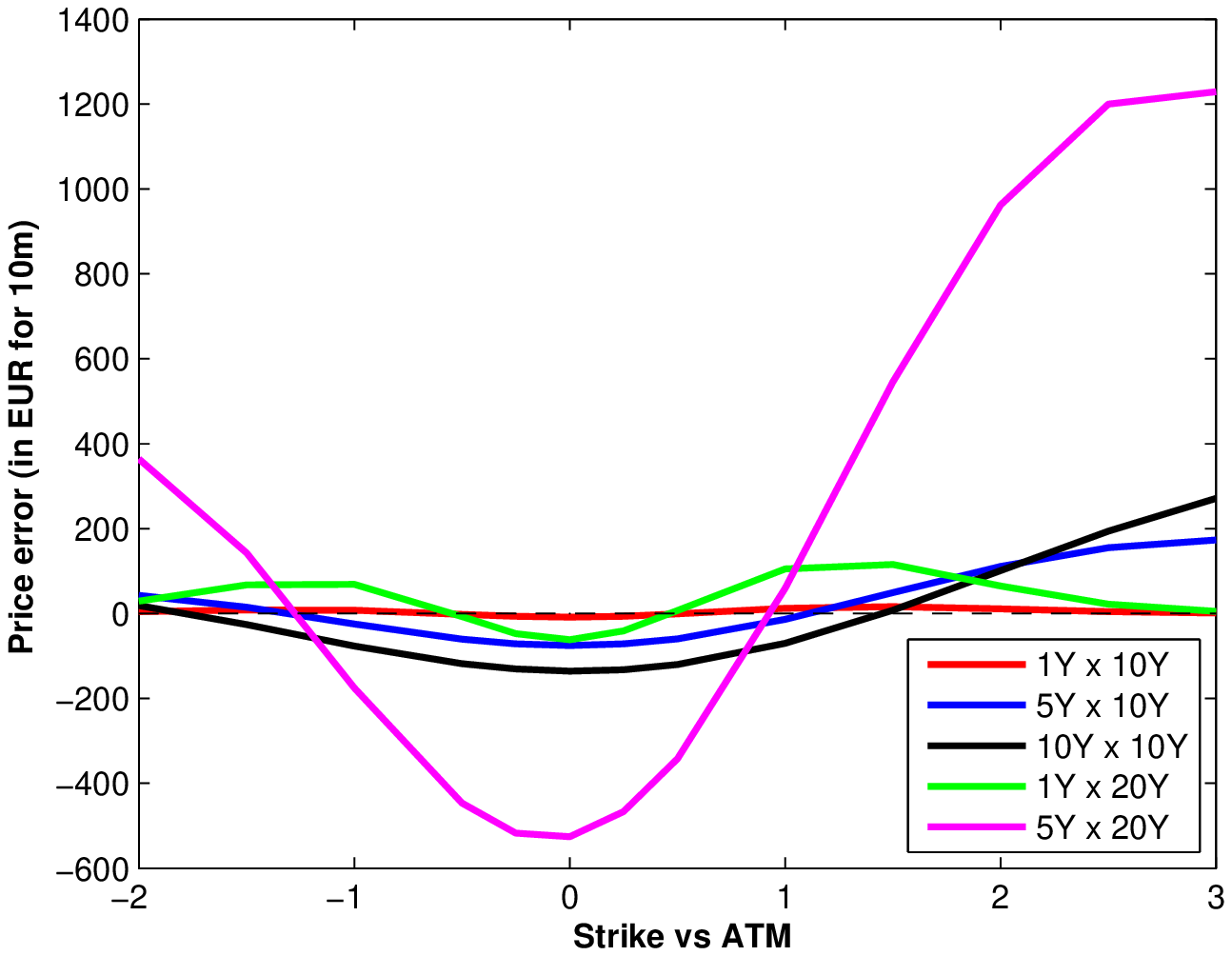, height=0.30 \textheight,width=0.48 \textwidth}
\label{FigErr2P} }
\end{center}
\caption{Error introduced by the approximation. Market data as of 19 decembre 2008.}
\label{FigErr2}
\end{figure}

The first example showed that the error is larger for long tenor options. For that reason the second example is concentrated on those options. The results are reproduced in Figure~\ref{FigErr2}. Note that in the second example, the strike below the money are only up to 250 bps away from the money. As the rates are around 3\%, a 300bps shift would give a strike close to 0.

In Figure~\ref{FigErr1V}, the proposed approximation is compared to other approximation proposed in the literature. The first comparison is with the initial freeze proposed by \cite{BAV.2006.1} (in the context of the BMM). In the graph, the initial freeze is represented by the dashed lines. That approximation is efficient for short tenor options but relatively inefficient for long tenors (10, 20 years) and strikes more than 100 bps from the money. As described in \cite{HEN.WP.9} the proposed approximation is at least ten times more efficient.

The second comparison is with the low variance martingale (LVM) approximation proposed by \cite{SCP.2006.1}. The approximation is represented in the graph by the dotted-dashed lines. The quality of that approximation is generally below the initial freeze one. The error is above the 0.10\% limit as soon as one goes more than 100 bps away from the money. The figures reported here are in line with the one reported in Table~4.2 of the original paper. The error is generally more than 25 times larger that the one with the approximation proposed here.

The second example, with market data significantly different, confirm the result of the first test. Up to 10 years tenors, the proposed approximation is sufficient. For longer tenors and strikes very far away from the money (more than 200 bps), the error may be above the a priori very stringent limit of 0.025\% but still below the market precision of 0.10\%. Note also that at that level of strikes, the market is relatively illiquid and their is a lot more uncertainly in the market level than in the approximation.

\subsection{Speed}

The speed testing is done using the last swaption in the second test. The price is computed on the author computer using a Matlab implementation. The price is computed 1000 times and the time required for the computations is recorded for each of the three implementations (exact, corrector approximation, initial freeze approximation). The result is 9.73s in the first case, 0.62s in the second and 0.58 in the third. The corrector approximation is around 15 times faster than the exact formula. The difference is in the computation of the $\kappa$, which is obtained by solving a one-dimensional non-linear equation. This takes around 90\% of the time for the exact formula. The speed gain using the initial freeze is only 8\% with respect to the corrector. Given the error which is significantly larger, the initial freeze approach is not advisable for calibration. The LVM approximation takes around 50\% more time than the corrector approximation and its precision is far less. There is no reason to use it in the Hull-White one-factor context.

\section{Conclusion}

An approximation for the pricing of European swaption in the Hull-White/extended Vasicek one factor model is proposed. The approximation is based on a corrector style approach. The approximation is proved very efficient with an error below the market precision. The approximation quality decreases with the underlying swap tenor and the swaption moneyness. Only for very long tenors and very far away from the money options the approximation quality decreases to be noticeable but always inferior to market precision.

The time efficiency of the implementation with corrector approximation is largely superior to the one of the exact formula. In our implementation the speed is increased fifteen fold. When a large number of computations is required, like in calibration, this gain can become significant. When speed is required we suggest to use the corrector approximation implementation.

The approximation approach presented in this note was initially developed for multi-factor Libor Market Model. It is more generic than presented here. It can also be applied to multi-factor Gaussian HJM model like the \emph{G2++} model.

\medskip

\textbf{Acknowledgment:} This note is dedicated to an efficient person who's birthdate is on December 20th, the date this note was started.

\medskip

\textbf{Disclaimer:} The views expressed here are those of the author and 
not necessarily those of its employers.

\bibliography{finance}

\begin{thebibliography}{}

\bibitem[\protect\astroncite{Baviera}{2006}]{BAV.2006.1}
Baviera, R. (2006).
\newblock Bond market model.
\newblock {\em International Journal of Theoretical and Applied Finance},
  9(4):577--596.

\bibitem[\protect\astroncite{Heath et~al.}{1992}]{HJM.1992.1}
Heath, D., Jarrow, R., and Morton, A. (1992).
\newblock Bond pricing and the term structure of interest rates: a new
  methodology for contingent claims valuation.
\newblock {\em Econometrica}, 60(1):77--105.

\bibitem[\protect\astroncite{Henrard}{2003}]{HEN.2003.1}
Henrard, M. (2003).
\newblock Explicit bond option and swaption formula in
  {H}eath-{J}arrow-{M}orton one-factor model.
\newblock {\em International Journal of Theoretical and Applied Finance},
  6(1):57--72.

\bibitem[\protect\astroncite{Henrard}{2005}]{HEN.2005.1}
Henrard, M. (2005).
\newblock Swaptions: 1 price, 10 deltas, and \ldots 6 1/2 gammas.
\newblock {\em Wilmott Magazine}, pages 48--57.

\bibitem[\protect\astroncite{Henrard}{2006}]{HEN.2006.1}
Henrard, M. (2006).
\newblock A semi-explicit approach to {C}anary swaptions in {HJM} one-factor
  model.
\newblock {\em Applied Mathematical Finance}, 13(1):1--18.
\newblock Preprint available at IDEAS:
  http://ideas.repec.org/p/wpa/wuwpfi/0310008.html.

\bibitem[\protect\astroncite{Henrard}{2007}]{HEN.2007.1}
Henrard, M. (2007).
\newblock Skewed {L}ibor {M}arket {M}odel and {G}aussian {HJM} explicit
  approaches to rolled deposit options.
\newblock {\em The Journal of Risk}, 9(4).
\newblock Preprint available at SSRN: http://ssrn.com/abstract=956849.

\bibitem[\protect\astroncite{Henrard}{2008}]{HEN.WP.9}
Henrard, M. (2008).
\newblock Swaptions in {L}ibor {M}arket {M}odel with local volatility.
\newblock Working paper, ???

\bibitem[\protect\astroncite{Hull and White}{1990}]{HUW.1990.1}
Hull, J. and White, A. (1990).
\newblock Pricing interest rate derivatives securities.
\newblock {\em The Review of Financial Studies}, 3:573--592.

\bibitem[\protect\astroncite{Hunt and Kennedy}{2004}]{HUK.2004.1}
Hunt, P.~J. and Kennedy, J.~E. (2004).
\newblock {\em Financial Derivatives in Theory and Practice}.
\newblock Wiley series in probability and statistics. Wiley, second edition.

\bibitem[\protect\astroncite{Hunter et~al.}{2001}]{HJJ.2001.1}
Hunter, C., J\"ackel, P., and Joshi, M. (2001).
\newblock Getting the drift.
\newblock {\em Risk}.

\bibitem[\protect\astroncite{Jamshidian}{1989}]{JAM.1989.1}
Jamshidian, F. (1989).
\newblock An exact bond option formula.
\newblock {\em The journal of Finance}, {XLIV}(1):205--209.

\bibitem[\protect\astroncite{Kl\"oden and Platen}{1995}]{KLP.1995.1}
Kl\"oden, P. and Platen, E. (1995).
\newblock {\em Numerical Solution of Stochastic Differential Equations}.
\newblock Spinger, Berlin, Heidelberg, New-York.

\bibitem[\protect\astroncite{Schrager and Pelsser}{2006}]{SCP.2006.1}
Schrager, D.~F. and Pelsser, A. A.~J. (2006).
\newblock Pricing swaptions and coupon bond options in affine term strucutre
  models.
\newblock {\em Mathematical Finance}, 16(4):673--694.

\end{thebibliography}
\bibliographystyle{apa}

\tableofcontents

\end{document}